\pdfoutput=1
\documentclass{article}

\usepackage[final]{ml4mols_2020}

\usepackage[utf8]{inputenc} 
\usepackage[T1]{fontenc}    
\usepackage{hyperref}       
\usepackage{url}            
\usepackage{booktabs}       
\usepackage{amsfonts}       
\usepackage{nicefrac}       
\usepackage{microtype}      
\usepackage{alphabeta}
\usepackage{amsmath,amsfonts,amssymb}
\usepackage{graphicx}
\usepackage{float}
\usepackage{wrapfig}
\usepackage{caption}
\usepackage{subcaption}
\usepackage{lipsum}  
\usepackage{amsmath}
\usepackage{authblk}
\title{Towards explainable message passing networks for predicting carbon dioxide adsorption in metal-organic frameworks}

\author[1]{\textbf{Ali Raza}\thanks{\texttt{razaa@oregonstate.edu}}~~}
\author[1]{\textbf{Faaiq Waqar}~}
\author[2]{\textbf{Arni Sturluson}~}
\author[2]{\textbf{Cory Simon} \thanks{\texttt{cory.simon@oregonstate.edu}}~~}
\author[1]{\textbf{Xiaoli Fern} \thanks{\texttt{xfern@oregonstate.edu}}~~}
\affil[1]{School of Electrical Engineering and Computer Science, Oregon State University}
\affil[2]{School of Chemical, Biological, and Environmental Engineering, 
   Oregon State University}

\begin{document}

\maketitle

\begin{abstract}
Metal-organic framework (MOFs) are nanoporous materials that could be used to capture carbon dioxide from the exhaust gas of fossil fuel power plants to mitigate climate change.
In this work, we design and train a message passing neural network (MPNN) to predict simulated CO$_2$ adsorption in MOFs.
Towards providing insights into what substructures of the MOFs are important for the prediction, we introduce a soft attention mechanism into the readout function that quantifies the contributions of the node representations towards the graph representations.
We investigate different mechanisms for sparse attention to ensure only the most relevant substructures are identified.

\end{abstract}

\section{Introduction}
\label{sec:introduction}
Anthropogenic carbon dioxide (CO$_2$) emissions are a major contributor to climate change and ocean acidification \cite{smit2014introduction}. Carbon dioxide capture and storage \cite{haszeldine2009carbon} is among a concerted portfolio of approaches \cite{pacala2004stabilization} to stabilize and eventually reduce our CO$_2$ emissions. 
In post-combustion carbon capture, CO$_2$ is separated from the combustion exhaust gas of fossil fuel power plants, at the point of emission, and then geologically sequestered \cite{smit2014introduction}.
Metal-organic frameworks (MOFs) \cite{furukawa2013chemistry} are nano-porous, crystalline materials that can selectively adsorb CO$_2$ \cite{sumida2012carbon,d2010carbon} and therefore could be used to capture CO$_2$ from the flue gas of fossil fuel power plants \cite{schoedel2016role}. 

MOFs are acclaimed as ``designer materials'' \cite{hendon2017grand} because the chemistry of the internal surface of the MOF can be (computationally) designed to target the adsorption of CO$_2$ \cite{boyd2019data}.
MOFs are synthesized modularly, by linking organic molecules to metals/metal clusters to form an extended network. 
Due to the abundance of molecular building blocks and their post-synthetic modifiability, the space of MOFs is vast.
Molecular models and simulations \cite{sturluson2019role,boyd2017computational} and machine learning \cite{chong2020applications,jablonka2020big,shi2020machine} play an important role in navigating this vast space of MOFs to find a suitable/optimal MOF for energy-efficient CO$_2$ capture and release \cite{huck2014evaluating}.

Here, we design and train a message passing neural network (MPNN) \cite{wu2019comprehensive,gilmer2017neural} to predict the (simulated) amount of CO$_2$ adsorption in MOFs.
As opposed to the traditional machine learning approach of human-engineering a feature vector to represent the structure of the MOF \cite{simon2015best,pardakhti2017machine,anderson2018role,dureckova2019robust,bucior2019energy}, the MPNN directly takes a graph representation of the MOF structure as input and automatically learns a vector representation of the MOF to use for the prediction task, in an end-to-end manner. This is achieved by iteratively passing information between neighboring nodes to learn hidden representations of the local bonding environments within the MOFs, then, through a readout function, aggregating the local node representations into a graph representation used for the prediction task. 
In a step towards explaining the predictions of the MPNN by identifying important substructures in the graph, we incorporated an attention mechanism in the readout function of the MPNN that quantifies the contribution of each node's representation to the graph representation.
Explainability is advantageous because it 
(i) can elucidate design rules and chemical intuition for synthesizing MOFs with desirable adsorption properties and
(ii) build appropriate trust/skepticism of particular predictions based on the explanation.

MPNNs \cite{wu2019comprehensive,gilmer2017neural} have been used to predict the properties of molecules and materials \cite{yang2019analyzing,kearnes2016molecular,stokes2020deep,gilmer2017neural,duvenaud2015convolutional,coley2017convolutional,jorgensen2018neural,tang2020self,sanchez2019machine,chen2019graph,st2019message,xie2018hierarchical,xie2018crystal,park2019developing}, as well as to generate molecules and materials with desired properties \cite{rocio,li2018learning}. There has been limited efforts in interpreting/explaining MPNNs or graph neural networks (GNNs) in general \cite{tang2020self, lin2020graph, huang2020graphlime, ji2020perturb, li2020understanding, ying2019gnnexplainer}.

\section{Proposed framework}
\subsection{Problem overview}
\label{sec:problem_formulation}
We aim to predict the equilibrium CO$_2$ adsorption in a MOF at a given temperature and pressure, $a \in \mathbb{R}^+$ [mmol/g]. 
Each MOF structure is represented as an undirected, node-labeled graph $G = (\mathcal{V}, \mathcal{E}, \mathbf{X})$, where $\mathcal{V}$ is the set of $n=|\mathcal{V}|$ nodes or vertices, representing atoms, $\mathcal{E}$ is the set of edges, representing bonds, and $\mathbf{X} \in \mathbb{R}^{d \times n}$ is the node feature matrix, whose columns are one-hot encodings of the chemical elements of the atoms ($d$ possible elements).
In a supervised manner, we aim to learn a function that maps a MOF to its predicted CO$_2$ adsorption:
$f: G \mapsto  f(G) =a$.

\begin{figure}[!t]
    \centering
    \includegraphics[width=\textwidth]{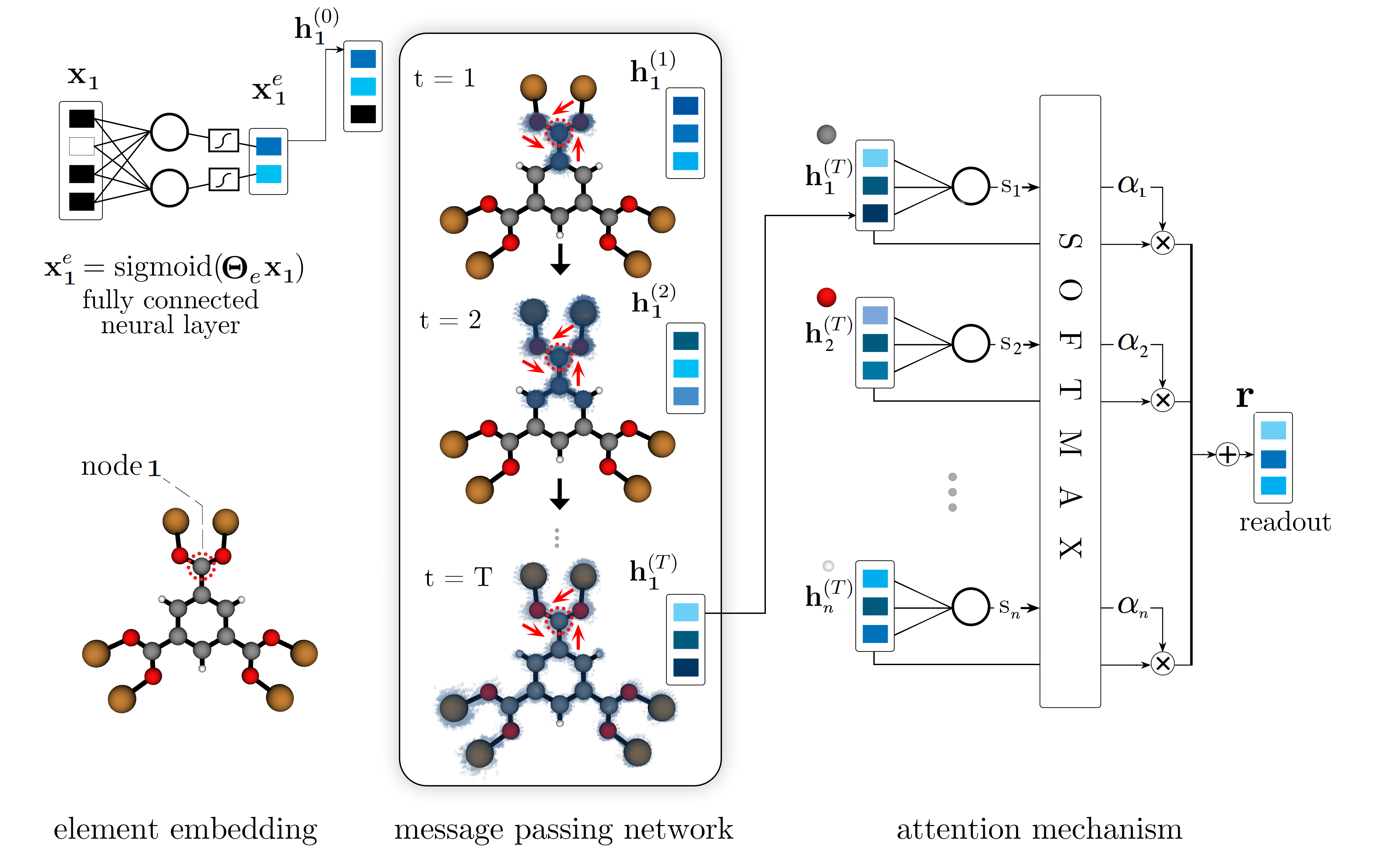}
    \caption{The architecture of our message passing neural network (MPNN).}
    \label{fig:architecutre}
\end{figure}

\subsection{Converting a MOF crystal structure to a graph}
We constructed the node-labeled graph $G$ representing each MOF from its unit cell, the list of atoms in the unit cell, and the crystallographic coordinates of those atoms. 
A bijection exists between the nodes $\mathcal{V}$ and the atoms comprising the unit cell of the MOF.
Two atoms are assigned an edge (bond) iff, as in Refs.~\cite{xie2018crystal,isayev2017universal,raza2020message}, 
(i) they are less than a distance $r$ apart, with $r$ the sum of their covalent radii \cite{cordero2008covalent} (some metals modified) plus a 0.25 {\AA} tolerance, 
and 
(ii) they share a Voronoi face in a Voronoi diagram of the surrounding atoms.
We used periodic distance in our calculations to include bonds across the periodic boundary of the unit cell.

\subsection{Message passing neural network that represents \texorpdfstring{$f$}{f}}
Fig.~\ref{fig:architecutre} shows the architecture of our message passing neural network (MPNN) that includes an attention mechanism to construct the graph representation from the set of node representations.

\paragraph{Message passing.}
First, our MPNN operates on nodes and learns a vector representation of the local bonding environment of each node in the graph. This is achieved by a chemical element embedding followed by iterations of message passing.

The chemical element embedding layer maps the one-hot encoding of the chemical element of the node to a low-dimensional, dense representation:
\begin{equation}
    \mathbf{x}^e_v = \texttt{sigmoid}.(\boldsymbol \Theta_e \mathbf{x}_v),
    \label{eq:embedding}
\end{equation} where $\mathbf{x}_v$ is column $v$ of $\mathbf{X}$ and $\boldsymbol \Theta_e$ is a $r \times d$ learned matrix, with $r<d$. 

Message passing is then used to learn a representation of each node encapsulating information about its local bonding environment.
Let $\mathbf{h}_v^{(t)}\in\mathbb{R}^k$ be the hidden representation of node $v$ at time step $t$, with $k \geq r$, initialized with the element embedding $\mathbf{x}^e_v$ padded with zeros. 
In each time step, every node receives information from its neighbors and updates its hidden representation accordingly, from an aggregated message, using a gated graph neural network (GGNN)~\cite{li2015gated}.
The aggregated message received by node $v$ is:
\begin{equation}
    \mathbf{m}_v^{(t+1)} = \boldsymbol  \Theta_m  \sum _{u \in \mathcal{N} (v)}  \mathbf{h}_{u}^{(t)}
    \label{eq:message}
\end{equation}
where $\boldsymbol \Theta_m$ is a learned $k \times k$ matrix shared across all nodes and $\mathcal{N} (v)$ is the set of nodes that share an edge with node $v$.
A Gated Recurrent Unit (GRU) (shared across all nodes) then updates the hidden representation of node $v$:
\begin{equation}
    \mathbf{h}_v^{(t+1)} = \texttt{GRU}(\mathbf{h}_v^{(t)}, \mathbf{m}_v^{(t+1)}).
\end{equation}
We conduct $T$ time steps of message passing, after which $\mathbf{h}_v^{(T)}$ contains information about the bonding environment of node $v$ within a graph-distance of $T$.

\paragraph{Readout and prediction.}
A readout function \cite{wu2020comprehensive} maps the set of hidden node representations to a fixed-size vector representation of the entire graph.
We use a soft attention mechanism~\cite{li2015gated} where the attention of node $v$, $\alpha_v$, is computed from the hidden features via a \texttt{softmax}:
\begin{equation}
    \alpha_v =   \frac{e^{\boldsymbol \theta_s^\intercal \mathbf{h}_v^{(T)}}}{\sum_{u=1}^n e^{\boldsymbol \theta_s^\intercal \mathbf{h}_u^{(T)}}}
\end{equation}
where $\boldsymbol \theta_s \in \mathbb{R}^k$ is a learned vector shared across all nodes.
To introduce sparsity, we also try (1) the quasi-norm L(0.5) regularization of the attention scores and (2) \texttt{sparsemax} \cite{martins2016softmax} in place of \texttt{softmax}.
The attention score of node $v$ then determines the contribution of its hidden representation $\mathbf{h}_v^{(T)}$ to the representation of the graph, $\mathbf{r}$:
\begin{equation}
\label{eq:prediction}
    \mathbf{r} = \sum_{v=1}^{n}  \alpha_v    \mathbf{h}_v^{(T)}.
\end{equation}

Finally, a neural network predicts the CO$_2$ adsorption ($\hat{a}$) from the graph representation:
\begin{equation}
    \hat{a} =   \texttt{softplus} \left( \boldsymbol \theta_{a}^\intercal \left( \texttt{sigmoid}. \left ( \boldsymbol \Theta_{a} \mathbf{r}  \right) \right) \right)
\end{equation}
where $\boldsymbol \Theta_{a}$ is a learned $z \times k$ matrix and $\boldsymbol \theta_{a} \in \mathbb{R}^z$ is a learned vector. The \texttt{softplus} ensures $a>0$.

\section{Results}
\label{sec:results}
As train, test, and validation data, we use simulated CO$_2$ uptake at 298\ K and 0.15\ bar from Ref.~\cite{moosavi2020understanding}, taken from the Materials Cloud \cite{talirz2020materials}, in 6\ 103 computation-ready, experimental MOF structures \cite{chung2019advances}.

\begin{table*}[!t]
\begin{center}
 \begin{tabular}{l l l l l} 
  \toprule
  \multicolumn{3}{r}{mean (std)}\\
  \cmidrule(r){2-4}
\textbf{Method} &\textbf{MAD}  &\textbf{MSE} & $\rho_r$ & Entropy (sparsity) \\
 \hline
 MPNN (softmax) & $0.616~(0.03)$ & $0.868~(0.10)$ & $0.764~(0.02)$ & $0.78$ ($0\%$) \\
 MPNN (sparsemax) & $0.666~(0.04)$ & $1.000~(0.11)$ & $0.732~(0.02)$ & $0.45$ ($94\%$) \\
 MPNN (L$0.5, \lambda = 0.001$) & $0.645~(0.02)$ & $0.933~(0.08)$ & $0.743~(0.01)$ & $0.72$ ($0\%$) \\
 MPNN (L$0.5, \lambda = 0.05$) & $0.737~(0.04)$ & $0.1.174~(0.13)$ & $0.684~(0.04)$ & $0.32$ ($0\%$) \\
 \bottomrule
\end{tabular}
\end{center}
\caption{Prediction performance and attention sparsity by different methods. Mean and standard deviation (std) over 10 folds.}
\label{table:quantitative_results}
\end{table*}
We use the mean absolute deviation (MAD) loss function $\ell=  \frac{1}{M} \sum_{m=1}^M ||\hat{{a}}_m - {a}_m||_1$ to train our network within $K=10$-fold cross validation,
where $M$ is the total number of MOFs, $\hat{{a}}_m$ is the predicted CO$_2$ adsorption of MOF $m$ predicted by the MPNN by eqn.~\ref{eq:prediction}, ${a}_m$ is the simulated CO$_2$ adsorption (treated as ground truth) and $|| \cdot ||_1$ is the L1 norm. Through hyperparameter exploration, we settled on $r=10$, $k=70$, and $T = 4$.
Tab.~\ref{table:quantitative_results} summarizes the performance of our model using the mean absolute deviation (MAD), mean square error (MSE), Spearman's rank correlation coefficient, $\rho_r$, and normalized entropy of the attentions (1 for uniform attention across all nodes and 0 for all attention concentrating on one node).
Vanilla \texttt{softmax} is able to achieve the best MAD performance. Fig.~\ref{fig:parity_plot} shows a parity plot, using softmax, for the test MOFs during the cross-validation procedure. Sparsemax introduced substantial sparsity in the attention scores ($\%94$ of the attention scores are zero); however, there is no way to control the sparsity. 
Using Quasi-norm L(0.5) regularisation produces a less uniform attention distribution, but with a price of slightly higher MAD. 
The regularization parameter, $\lambda$, enables us to trade-off training set accuracy with sparsity. 
Smaller entropy ($\lambda = 0.001$ to $0.05)$ results in less accuracy (MAD = $0.645$ to $0.737$).

\section{Discussion: towards explainability}
\label{discussion}

\begin{figure}[!h]
    \begin{subfigure}{0.39\textwidth}
        \includegraphics[width=\linewidth]{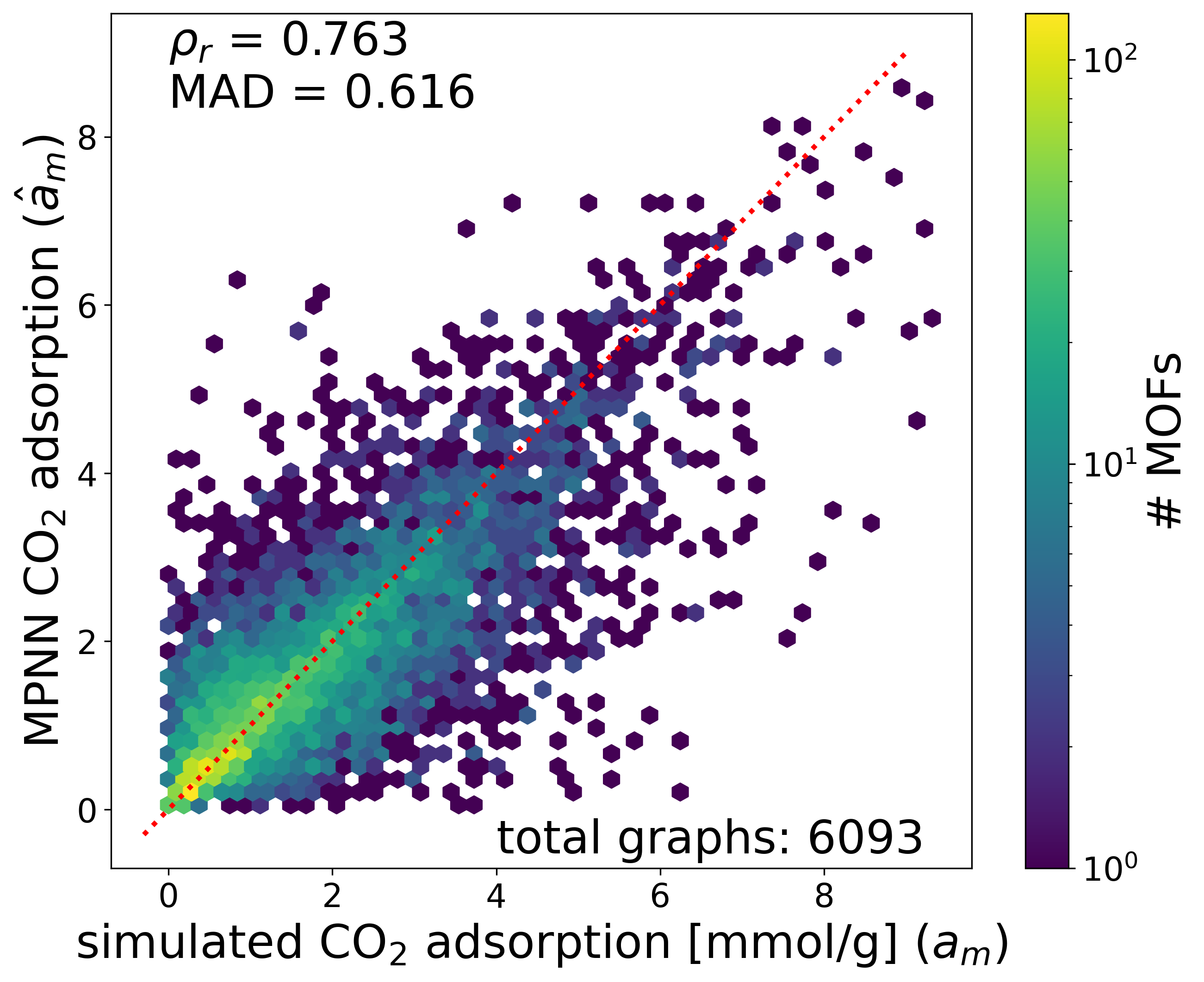}
        \caption{}
        \label{fig:parity_plot}
    \end{subfigure}
    \begin{subfigure}{0.29\textwidth}
        \includegraphics[width=\linewidth]{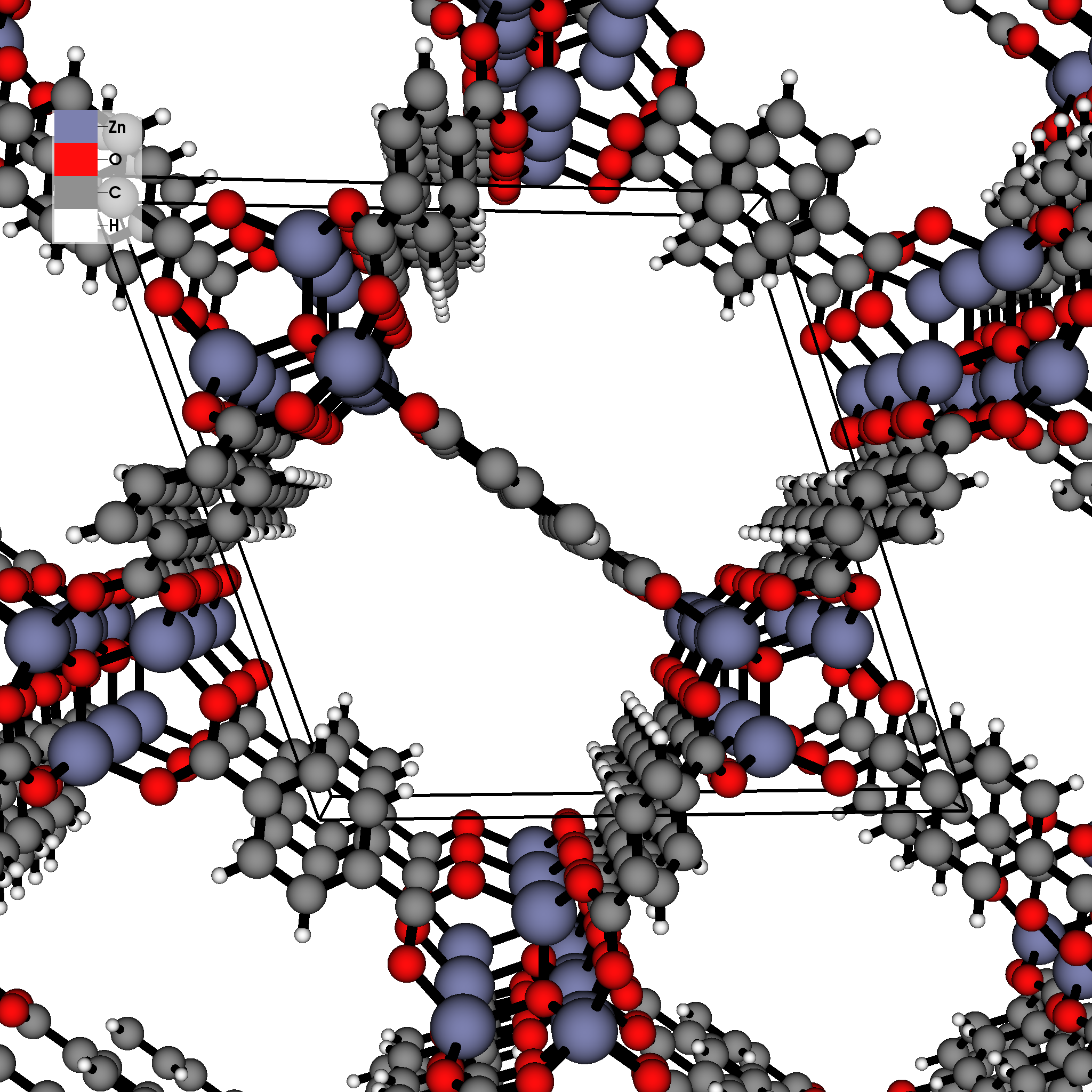} 
        \caption{}
        \label{fig:IRMOF1_elements}
        \end{subfigure}
    \begin{subfigure}{0.29\textwidth}
        \includegraphics[width=\linewidth]{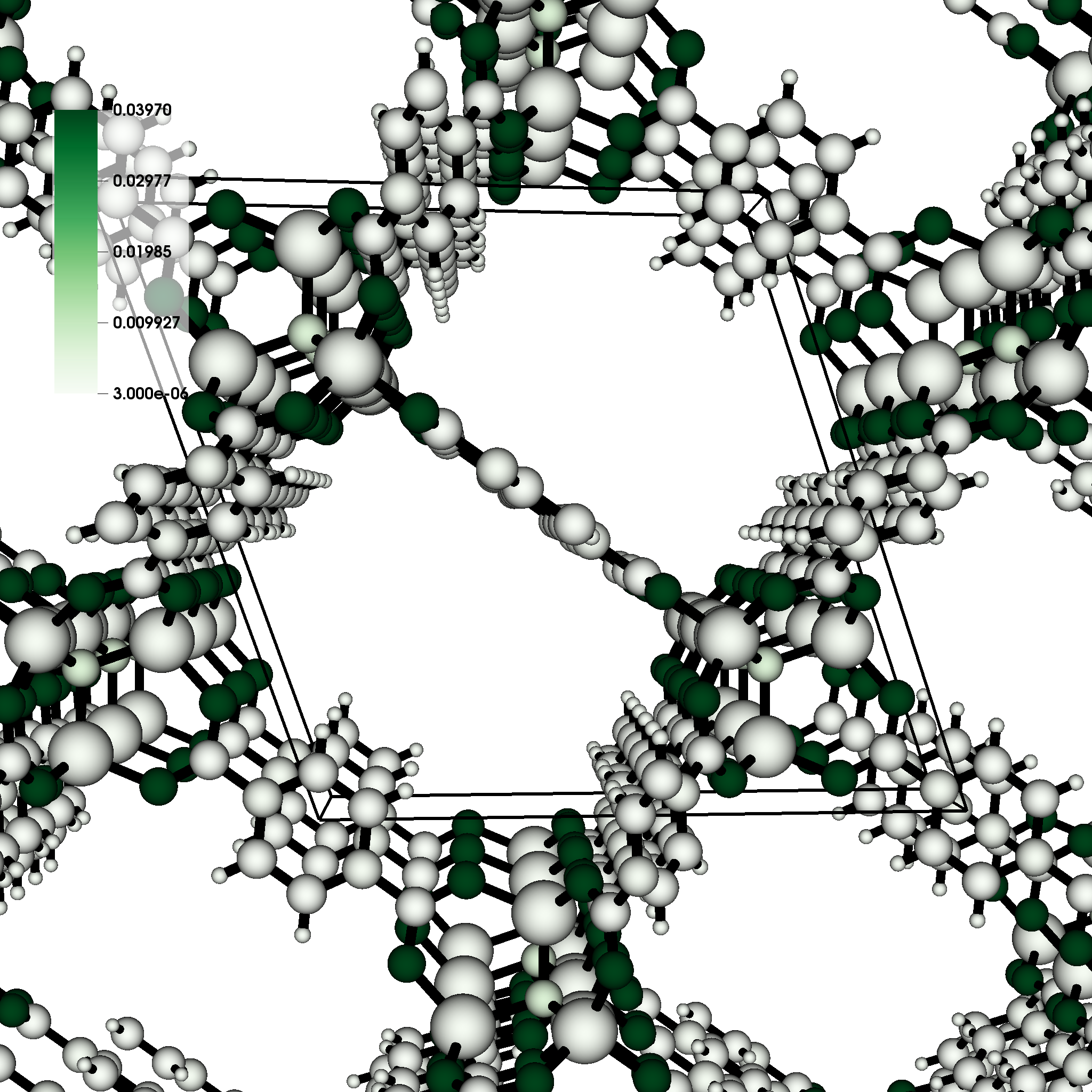}
        \caption{}
        \label{fig:IRMOF1_attention}
    \end{subfigure}
    \caption{(a) Parity plot for the MPNN (softmax), showing the MPNN-predicted vs.\ simulated CO$_2$ adsorption, including every MOF in the data set when it served as a test MOF in cross validation. Diagonal line shows equality. (b, c)
    Visualization of MOF IRMOF-1, where color indicates (b) the identity of the atom and (c) the attention score $\alpha_v$ on the atom using the MPNN (softmax).
    Black boxes $=$ unit cell. 
}
    \label{fig:image2}
\end{figure}

We include the attention score $\alpha_v$ as a step towards an MPNN with explainable predictions. If $\alpha_v$ is large, the hidden representation of node $v$ had a significant contribution to the graph representation $\mathbf{r}$ used to predict adsorption, $a$. 
Fig.~\ref{fig:IRMOF1_attention} visualizes the attention of each node in a MOF as an example. The local bonding environments of the darker atoms contributed more to the final graph representation used for the prediction task than the lighter atoms.

Our inspection of the attention scores across different MOFs did not yield any clear insight or chemically meaningful patterns. This leads us to believe that the attention score is not as meaningful as we have hoped for identifying important substructures. we further believe this is due to (i) $\mathbf{h}_v^{(T)}$ of node $v$ is enriched with information from all nodes within $T$ steps from node $v$ via message passing; (ii) neighboring nodes tend to have strong similarity; and (iii) the attention scores do not necessarily reflect the importance of node $v$ for the predicted adsorption \cite{jain2019attention}. Therefore, we are currently working to redesign the MPNN to short-circuit message passing, isolate the individual contributions of the nodes, and rigorously lend more explainability of the prediction.


\begin{ack}
The authors acknowledge the National Science Foundation for support under grants No.\ 1920945 and No.\ 1521687.
\end{ack}

\bibliographystyle{unsrt}
\bibliography{}

\begin{thebibliography}{52}
\providecommand{\natexlab}[1]{#1}
\providecommand{\url}[1]{\texttt{#1}}
\expandafter\ifx\csname urlstyle\endcsname\relax
  \providecommand{\doi}[1]{doi: #1}\else
  \providecommand{\doi}{doi: \begingroup \urlstyle{rm}\Url}\fi

\bibitem[Anderson et~al.(2018)Anderson, Rodgers, Argueta, Biong, and
  G{\'o}mez-Gualdr{\'o}n]{anderson2018role}
Ryther Anderson, Jacob Rodgers, Edwin Argueta, Achay Biong, and Diego~A
  G{\'o}mez-Gualdr{\'o}n.
\newblock Role of pore chemistry and topology in the co2 capture capabilities
  of mofs: from molecular simulation to machine learning.
\newblock \emph{Chemistry of Materials}, 30\penalty0 (18):\penalty0 6325--6337,
  2018.

\bibitem[Boyd et~al.(2017)Boyd, Lee, and Smit]{boyd2017computational}
Peter~G Boyd, Yongjin Lee, and Berend Smit.
\newblock Computational development of the nanoporous materials genome.
\newblock \emph{Nature Reviews Materials}, 2\penalty0 (8):\penalty0 1--15,
  2017.

\bibitem[Boyd et~al.(2019)Boyd, Chidambaram, Garc{\'\i}a-D{\'\i}ez, Ireland,
  Daff, Bounds, G{\l}adysiak, Schouwink, Moosavi, Maroto-Valer,
  et~al.]{boyd2019data}
Peter~G Boyd, Arunraj Chidambaram, Enrique Garc{\'\i}a-D{\'\i}ez, Christopher~P
  Ireland, Thomas~D Daff, Richard Bounds, Andrzej G{\l}adysiak, Pascal
  Schouwink, Seyed~Mohamad Moosavi, M~Mercedes Maroto-Valer, et~al.
\newblock Data-driven design of metal--organic frameworks for wet flue gas co 2
  capture.
\newblock \emph{Nature}, 576\penalty0 (7786):\penalty0 253--256, 2019.

\bibitem[Bucior et~al.(2019)Bucior, Bobbitt, Islamoglu, Goswami, Gopalan,
  Yildirim, Farha, Bagheri, and Snurr]{bucior2019energy}
Benjamin~J Bucior, N~Scott Bobbitt, Timur Islamoglu, Subhadip Goswami, Arun
  Gopalan, Taner Yildirim, Omar~K Farha, Neda Bagheri, and Randall~Q Snurr.
\newblock Energy-based descriptors to rapidly predict hydrogen storage in
  metal--organic frameworks.
\newblock \emph{Molecular Systems Design \& Engineering}, 4\penalty0
  (1):\penalty0 162--174, 2019.

\bibitem[Chen et~al.(2019)Chen, Ye, Zuo, Zheng, and Ong]{chen2019graph}
Chi Chen, Weike Ye, Yunxing Zuo, Chen Zheng, and Shyue~Ping Ong.
\newblock Graph networks as a universal machine learning framework for
  molecules and crystals.
\newblock \emph{Chemistry of Materials}, 31\penalty0 (9):\penalty0 3564--3572,
  2019.

\bibitem[Chong et~al.(2020)Chong, Lee, Kim, and Kim]{chong2020applications}
Sanggyu Chong, Sangwon Lee, Baekjun Kim, and Jihan Kim.
\newblock Applications of machine learning in metal-organic frameworks.
\newblock \emph{Coordination Chemistry Reviews}, 423:\penalty0 213487, 2020.

\bibitem[Chung et~al.(2019)Chung, Haldoupis, Bucior, Haranczyk, Lee, Zhang,
  Vogiatzis, Milisavljevic, Ling, Camp, et~al.]{chung2019advances}
Yongchul~G Chung, Emmanuel Haldoupis, Benjamin~J Bucior, Maciej Haranczyk,
  Seulchan Lee, Hongda Zhang, Konstantinos~D Vogiatzis, Marija Milisavljevic,
  Sanliang Ling, Jeffrey~S Camp, et~al.
\newblock Advances, updates, and analytics for the computation-ready,
  experimental metal--organic framework database: Core mof 2019.
\newblock \emph{Journal of Chemical \& Engineering Data}, 64\penalty0
  (12):\penalty0 5985--5998, 2019.

\bibitem[Coley et~al.(2017)Coley, Barzilay, Green, Jaakkola, and
  Jensen]{coley2017convolutional}
Connor~W Coley, Regina Barzilay, William~H Green, Tommi~S Jaakkola, and Klavs~F
  Jensen.
\newblock Convolutional embedding of attributed molecular graphs for physical
  property prediction.
\newblock \emph{Journal of Chemical Information and Modeling}, 57\penalty0
  (8):\penalty0 1757--1772, 2017.

\bibitem[Cordero et~al.(2008)Cordero, G{\'o}mez, Platero-Prats, Rev{\'e}s,
  Echeverr{\'\i}a, Cremades, Barrag{\'a}n, and Alvarez]{cordero2008covalent}
Beatriz Cordero, Ver{\'o}nica G{\'o}mez, Ana~E Platero-Prats, Marc Rev{\'e}s,
  Jorge Echeverr{\'\i}a, Eduard Cremades, Flavia Barrag{\'a}n, and Santiago
  Alvarez.
\newblock Covalent radii revisited.
\newblock \emph{Dalton Transactions}, \penalty0 (21):\penalty0 2832--2838,
  2008.

\bibitem[D'Alessandro et~al.(2010)D'Alessandro, Smit, and Long]{d2010carbon}
Deanna~M D'Alessandro, Berend Smit, and Jeffrey~R Long.
\newblock Carbon dioxide capture: prospects for new materials.
\newblock \emph{Angewandte Chemie International Edition}, 49\penalty0
  (35):\penalty0 6058--6082, 2010.

\bibitem[Dureckova et~al.(2019)Dureckova, Krykunov, Aghaji, and
  Woo]{dureckova2019robust}
Hana Dureckova, Mykhaylo Krykunov, Mohammad~Zein Aghaji, and Tom~K Woo.
\newblock Robust machine learning models for predicting high co2 working
  capacity and co2/h2 selectivity of gas adsorption in metal organic frameworks
  for precombustion carbon capture.
\newblock \emph{The Journal of Physical Chemistry C}, 123\penalty0
  (7):\penalty0 4133--4139, 2019.

\bibitem[Duvenaud et~al.(2015)Duvenaud, Maclaurin, Iparraguirre, Bombarell,
  Hirzel, Aspuru-Guzik, and Adams]{duvenaud2015convolutional}
David~K Duvenaud, Dougal Maclaurin, Jorge Iparraguirre, Rafael Bombarell,
  Timothy Hirzel, Al{\'a}n Aspuru-Guzik, and Ryan~P Adams.
\newblock Convolutional networks on graphs for learning molecular fingerprints.
\newblock In \emph{Advances in Neural Information Processing Systems}, pages
  2224--2232, 2015.

\bibitem[Furukawa et~al.(2013)Furukawa, Cordova, O’Keeffe, and
  Yaghi]{furukawa2013chemistry}
Hiroyasu Furukawa, Kyle~E Cordova, Michael O’Keeffe, and Omar~M Yaghi.
\newblock The chemistry and applications of metal-organic frameworks.
\newblock \emph{Science}, 341\penalty0 (6149):\penalty0 1230444, 2013.

\bibitem[Gilmer et~al.(2017)Gilmer, Schoenholz, Riley, Vinyals, and
  Dahl]{gilmer2017neural}
Justin Gilmer, Samuel~S Schoenholz, Patrick~F Riley, Oriol Vinyals, and
  George~E Dahl.
\newblock Neural message passing for quantum chemistry.
\newblock In \emph{Proceedings of the 34th International Conference on Machine
  Learning-Volume 70}, pages 1263--1272. JMLR. org, 2017.

\bibitem[Haszeldine(2009)]{haszeldine2009carbon}
R~Stuart Haszeldine.
\newblock Carbon capture and storage: how green can black be?
\newblock \emph{Science}, 325\penalty0 (5948):\penalty0 1647--1652, 2009.

\bibitem[Hendon et~al.(2017)Hendon, Rieth, Korzy{\'n}ski, and
  Dinc{\u{a}}]{hendon2017grand}
Christopher~H Hendon, Adam~J Rieth, Maciej~D Korzy{\'n}ski, and Mircea
  Dinc{\u{a}}.
\newblock Grand challenges and future opportunities for metal--organic
  frameworks.
\newblock \emph{ACS Central Science}, 3\penalty0 (6):\penalty0 554--563, 2017.

\bibitem[Huang et~al.(2020)Huang, Yamada, Tian, Singh, Yin, and
  Chang]{huang2020graphlime}
Qiang Huang, Makoto Yamada, Yuan Tian, Dinesh Singh, Dawei Yin, and Yi~Chang.
\newblock Graphlime: Local interpretable model explanations for graph neural
  networks.
\newblock \emph{arXiv preprint arXiv:2001.06216}, 2020.

\bibitem[Huck et~al.(2014)Huck, Lin, Berger, Shahrak, Martin, Bhown, Haranczyk,
  Reuter, and Smit]{huck2014evaluating}
Johanna~M Huck, Li-Chiang Lin, Adam~H Berger, Mahdi~Niknam Shahrak, Richard~L
  Martin, Abhoyjit~S Bhown, Maciej Haranczyk, Karsten Reuter, and Berend Smit.
\newblock Evaluating different classes of porous materials for carbon capture.
\newblock \emph{Energy \& Environmental Science}, 7\penalty0 (12):\penalty0
  4132--4146, 2014.

\bibitem[Isayev et~al.(2017)Isayev, Oses, Toher, Gossett, Curtarolo, and
  Tropsha]{isayev2017universal}
Olexandr Isayev, Corey Oses, Cormac Toher, Eric Gossett, Stefano Curtarolo, and
  Alexander Tropsha.
\newblock Universal fragment descriptors for predicting properties of inorganic
  crystals.
\newblock \emph{Nature Communications}, 8\penalty0 (1):\penalty0 1--12, 2017.

\bibitem[Jablonka et~al.(2020)Jablonka, Ongari, Moosavi, and
  Smit]{jablonka2020big}
Kevin~Maik Jablonka, Daniele Ongari, Seyed~Mohamad Moosavi, and Berend Smit.
\newblock Big-data science in porous materials: Materials genomics and machine
  learning.
\newblock \emph{Chemical Reviews}, 2020.

\bibitem[Jain and Wallace(2019)]{jain2019attention}
Sarthak Jain and Byron~C Wallace.
\newblock Attention is not explanation.
\newblock \emph{arXiv preprint arXiv:1902.10186}, 2019.

\bibitem[Ji et~al.(2020)Ji, Wang, Li, and Wu]{ji2020perturb}
Chaojie Ji, Ruxin Wang, Ye~Li, and Hongyan Wu.
\newblock Perturb more, trap more: Understanding behaviors of graph neural
  networks.
\newblock \emph{arXiv preprint arXiv:2004.09808}, 2020.

\bibitem[J{\o}rgensen et~al.(2018)J{\o}rgensen, Jacobsen, and
  Schmidt]{jorgensen2018neural}
Peter~Bj{\o}rn J{\o}rgensen, Karsten~Wedel Jacobsen, and Mikkel~N Schmidt.
\newblock Neural message passing with edge updates for predicting properties of
  molecules and materials.
\newblock \emph{arXiv preprint arXiv:1806.03146}, 2018.

\bibitem[Kearnes et~al.(2016)Kearnes, McCloskey, Berndl, Pande, and
  Riley]{kearnes2016molecular}
Steven Kearnes, Kevin McCloskey, Marc Berndl, Vijay Pande, and Patrick Riley.
\newblock Molecular graph convolutions: moving beyond fingerprints.
\newblock \emph{Journal of Computer-aided Molecular Design}, 30\penalty0
  (8):\penalty0 595--608, 2016.

\bibitem[Li and Cheng(2020)]{li2020understanding}
Xue Li and Yuanzhi Cheng.
\newblock Understanding the message passing in graph neural networks via power
  iteration.
\newblock \emph{arXiv preprint arXiv:2006.00144}, 2020.

\bibitem[Li et~al.(2015)Li, Tarlow, Brockschmidt, and Zemel]{li2015gated}
Yujia Li, Daniel Tarlow, Marc Brockschmidt, and Richard Zemel.
\newblock Gated graph sequence neural networks.
\newblock \emph{arXiv preprint arXiv:1511.05493}, 2015.

\bibitem[Li et~al.(2018)Li, Vinyals, Dyer, Pascanu, and
  Battaglia]{li2018learning}
Yujia Li, Oriol Vinyals, Chris Dyer, Razvan Pascanu, and Peter Battaglia.
\newblock Learning deep generative models of graphs.
\newblock \emph{arXiv preprint arXiv:1803.03324}, 2018.

\bibitem[Lin et~al.(2020)Lin, Sun, Bulusu, Dry, and Hernandez]{lin2020graph}
Chris Lin, Gerald~J Sun, Krishna~C Bulusu, Jonathan~R Dry, and Marylens
  Hernandez.
\newblock Graph neural networks including sparse interpretability.
\newblock \emph{arXiv preprint arXiv:2007.00119}, 2020.

\bibitem[Martins and Astudillo(2016)]{martins2016softmax}
Andre Martins and Ramon Astudillo.
\newblock From softmax to sparsemax: A sparse model of attention and
  multi-label classification.
\newblock In \emph{International Conference on Machine Learning}, pages
  1614--1623, 2016.

\bibitem[Mercado et~al.(2020)Mercado, Rastemo, Lindelöf, Klambauer, Engkvist,
  Chen, and Bjerrum]{rocio}
Rocío Mercado, Tobias Rastemo, Edvard Lindelöf, Günter Klambauer, Ola
  Engkvist, Hongming Chen, and Esben~Jannik Bjerrum.
\newblock Graph networks for molecular design.
\newblock \emph{ChemRxiv}, 2020.
\newblock \doi{10.26434/chemrxiv.12843137.v1}.
\newblock URL
  \url{https://chemrxiv.org/articles/preprint/Graph_Networks_for_Molecular_Design/12843137/1}.

\bibitem[Moosavi et~al.(2020)Moosavi, Nandy, Jablonka, Ongari, Janet, Boyd,
  Lee, Smit, and Kulik]{moosavi2020understanding}
Seyed~Mohamad Moosavi, Aditya Nandy, Kevin~Maik Jablonka, Daniele Ongari,
  Jon~Paul Janet, Peter~G Boyd, Yongjin Lee, Berend Smit, and Heather~J Kulik.
\newblock Understanding the diversity of the metal-organic framework ecosystem.
\newblock \emph{Nature Communications}, 11\penalty0 (1):\penalty0 1--10, 2020.

\bibitem[Pacala and Socolow(2004)]{pacala2004stabilization}
Stephen Pacala and Robert Socolow.
\newblock Stabilization wedges: solving the climate problem for the next 50
  years with current technologies.
\newblock \emph{Science}, 305\penalty0 (5686):\penalty0 968--972, 2004.

\bibitem[Pardakhti et~al.(2017)Pardakhti, Moharreri, Wanik, Suib, and
  Srivastava]{pardakhti2017machine}
Maryam Pardakhti, Ehsan Moharreri, David Wanik, Steven~L Suib, and Ranjan
  Srivastava.
\newblock Machine learning using combined structural and chemical descriptors
  for prediction of methane adsorption performance of metal organic frameworks
  (mofs).
\newblock \emph{ACS Combinatorial Science}, 19\penalty0 (10):\penalty0
  640--645, 2017.

\bibitem[Park and Wolverton(2019)]{park2019developing}
Cheol~Woo Park and Chris Wolverton.
\newblock Developing an improved crystal graph convolutional neural network
  framework for accelerated materials discovery.
\newblock \emph{arXiv preprint arXiv:1906.05267}, 2019.

\bibitem[Raza et~al.(2020)Raza, Sturluson, Simon, and Fern]{raza2020message}
Ali Raza, Arni Sturluson, Cory~M Simon, and Xiaoli Fern.
\newblock Message passing neural networks for partial charge assignment to
  metal--organic frameworks.
\newblock \emph{The Journal of Physical Chemistry C}, 124\penalty0
  (35):\penalty0 19070--19082, 2020.

\bibitem[Sanchez-Lengeling et~al.(2019)Sanchez-Lengeling, Wei, Lee, Gerkin,
  Aspuru-Guzik, and Wiltschko]{sanchez2019machine}
Benjamin Sanchez-Lengeling, Jennifer~N Wei, Brian~K Lee, Richard~C Gerkin,
  Al{\'a}n Aspuru-Guzik, and Alexander~B Wiltschko.
\newblock Machine learning for scent: Learning generalizable perceptual
  representations of small molecules.
\newblock \emph{arXiv preprint arXiv:1910.10685}, 2019.

\bibitem[Schoedel et~al.(2016)Schoedel, Ji, and Yaghi]{schoedel2016role}
Alexander Schoedel, Zhe Ji, and Omar~M Yaghi.
\newblock The role of metal--organic frameworks in a carbon-neutral energy
  cycle.
\newblock \emph{Nature Energy}, 1\penalty0 (4):\penalty0 1--13, 2016.

\bibitem[Shi et~al.(2020)Shi, Yang, Deng, Cai, Yan, Liang, Liu, and
  Qiao]{shi2020machine}
Zenan Shi, Wenyuan Yang, Xiaomei Deng, Chengzhi Cai, Yaling Yan, Hong Liang,
  Zili Liu, and Zhiwei Qiao.
\newblock Machine-learning-assisted high-throughput computational screening of
  high performance metal--organic frameworks.
\newblock \emph{Molecular Systems Design \& Engineering}, 5\penalty0
  (4):\penalty0 725--742, 2020.

\bibitem[Simon et~al.(2015)Simon, Mercado, Schnell, Smit, and
  Haranczyk]{simon2015best}
Cory~M Simon, Rocio Mercado, Sondre~K Schnell, Berend Smit, and Maciej
  Haranczyk.
\newblock What are the best materials to separate a xenon/krypton mixture?
\newblock \emph{Chemistry of Materials}, 27\penalty0 (12):\penalty0 4459--4475,
  2015.

\bibitem[Smit et~al.(2014)Smit, Reimer, Oldenburg, and
  Bourg]{smit2014introduction}
Berend Smit, Jeffrey~A Reimer, Curtis~M Oldenburg, and Ian~C Bourg.
\newblock \emph{Introduction to carbon capture and sequestration}, volume~1.
\newblock World Scientific, 2014.

\bibitem[St.~John et~al.(2019)St.~John, Phillips, Kemper, Wilson, Guan,
  Crowley, Nimlos, and Larsen]{st2019message}
Peter~C St.~John, Caleb Phillips, Travis~W Kemper, A~Nolan Wilson, Yanfei Guan,
  Michael~F Crowley, Mark~R Nimlos, and Ross~E Larsen.
\newblock Message-passing neural networks for high-throughput polymer
  screening.
\newblock \emph{The Journal of Chemical Physics}, 150\penalty0 (23):\penalty0
  234111, 2019.

\bibitem[Stokes et~al.(2020)Stokes, Yang, Swanson, Jin, Cubillos-Ruiz, Donghia,
  MacNair, French, Carfrae, Bloom-Ackerman, et~al.]{stokes2020deep}
Jonathan~M Stokes, Kevin Yang, Kyle Swanson, Wengong Jin, Andres Cubillos-Ruiz,
  Nina~M Donghia, Craig~R MacNair, Shawn French, Lindsey~A Carfrae, Zohar
  Bloom-Ackerman, et~al.
\newblock A deep learning approach to antibiotic discovery.
\newblock \emph{Cell}, 180\penalty0 (4):\penalty0 688--702, 2020.

\bibitem[Sturluson et~al.(2019)Sturluson, Huynh, Kaija, Laird, Yoon, Hou, Feng,
  Wilmer, Col{\'o}n, Chung, D, and C]{sturluson2019role}
Arni Sturluson, Melanie~T Huynh, Alec~R Kaija, Caleb Laird, Sunghyun Yoon,
  Feier Hou, Zhenxing Feng, Christopher~E Wilmer, Yamil~J Col{\'o}n, Yongchul~G
  Chung, Siderius D, and Simon C.
\newblock The role of molecular modelling and simulation in the discovery and
  deployment of metal-organic frameworks for gas storage and separation.
\newblock \emph{Molecular Simulation}, 45\penalty0 (14-15):\penalty0
  1082--1121, 2019.

\bibitem[Sumida et~al.(2012)Sumida, Rogow, Mason, McDonald, Bloch, Herm, Bae,
  and Long]{sumida2012carbon}
Kenji Sumida, David~L Rogow, Jarad~A Mason, Thomas~M McDonald, Eric~D Bloch,
  Zoey~R Herm, Tae-Hyun Bae, and Jeffrey~R Long.
\newblock Carbon dioxide capture in metal--organic frameworks.
\newblock \emph{Chemical Reviews}, 112\penalty0 (2):\penalty0 724--781, 2012.

\bibitem[Talirz et~al.(2020)Talirz, Kumbhar, Passaro, Yakutovich, Granata,
  Gargiulo, Borelli, Uhrin, Huber, Zoupanos, et~al.]{talirz2020materials}
Leopold Talirz, Snehal Kumbhar, Elsa Passaro, Aliaksandr~V Yakutovich, Valeria
  Granata, Fernando Gargiulo, Marco Borelli, Martin Uhrin, Sebastiaan~P Huber,
  Spyros Zoupanos, et~al.
\newblock Materials cloud, a platform for open computational science.
\newblock \emph{arXiv preprint arXiv:2003.12510}, 2020.

\bibitem[Tang et~al.(2020)Tang, Kramer, Fang, Qiu, Wu, and Xu]{tang2020self}
Bowen Tang, Skyler~T Kramer, Meijuan Fang, Yingkun Qiu, Zhen Wu, and Dong Xu.
\newblock A self-attention based message passing neural network for predicting
  molecular lipophilicity and aqueous solubility.
\newblock \emph{Journal of Cheminformatics}, 12\penalty0 (1):\penalty0 1--9,
  2020.

\bibitem[Wu et~al.(2019)Wu, Pan, Chen, Long, Zhang, and
  Yu]{wu2019comprehensive}
Zonghan Wu, Shirui Pan, Fengwen Chen, Guodong Long, Chengqi Zhang, and Philip~S
  Yu.
\newblock A comprehensive survey on graph neural networks.
\newblock \emph{arXiv preprint arXiv:1901.00596}, 2019.

\bibitem[Wu et~al.(2020)Wu, Pan, Chen, Long, Zhang, and
  Philip]{wu2020comprehensive}
Zonghan Wu, Shirui Pan, Fengwen Chen, Guodong Long, Chengqi Zhang, and S~Yu
  Philip.
\newblock A comprehensive survey on graph neural networks.
\newblock \emph{IEEE Transactions on Neural Networks and Learning Systems},
  2020.

\bibitem[Xie and Grossman(2018{\natexlab{a}})]{xie2018crystal}
Tian Xie and Jeffrey~C Grossman.
\newblock Crystal graph convolutional neural networks for an accurate and
  interpretable prediction of material properties.
\newblock \emph{Physical Review Letters}, 120\penalty0 (14):\penalty0 145301,
  2018{\natexlab{a}}.

\bibitem[Xie and Grossman(2018{\natexlab{b}})]{xie2018hierarchical}
Tian Xie and Jeffrey~C Grossman.
\newblock Hierarchical visualization of materials space with graph
  convolutional neural networks.
\newblock \emph{The Journal of Chemical Physics}, 149\penalty0 (17):\penalty0
  174111, 2018{\natexlab{b}}.

\bibitem[Yang et~al.(2019)Yang, Swanson, Jin, Coley, Eiden, Gao, Guzman-Perez,
  Hopper, Kelley, Mathea, et~al.]{yang2019analyzing}
Kevin Yang, Kyle Swanson, Wengong Jin, Connor Coley, Philipp Eiden, Hua Gao,
  Angel Guzman-Perez, Timothy Hopper, Brian Kelley, Miriam Mathea, et~al.
\newblock Analyzing learned molecular representations for property prediction.
\newblock \emph{Journal of Chemical Information and Modeling}, 59\penalty0
  (8):\penalty0 3370--3388, 2019.

\bibitem[Ying et~al.(2019)Ying, Bourgeois, You, Zitnik, and
  Leskovec]{ying2019gnnexplainer}
Zhitao Ying, Dylan Bourgeois, Jiaxuan You, Marinka Zitnik, and Jure Leskovec.
\newblock Gnnexplainer: Generating explanations for graph neural networks.
\newblock In \emph{Advances in neural information processing systems}, pages
  9244--9255, 2019.

\end{thebibliography}

\end{document}